\documentclass[prd,superscriptaddress]{revtex4}
\usepackage{CJK}
\usepackage{amsmath}
\usepackage{enumerate}
\usepackage{amssymb}
\usepackage{amsfonts}
\usepackage{mathrsfs}
\usepackage{latexsym}
\usepackage{bm}
\usepackage{amscd}
\usepackage{graphicx}
\usepackage{epsfig}
\usepackage{hhline,multirow}
\usepackage{dcolumn}
\usepackage{url}
\usepackage[colorlinks=true,linkcolor=blue]{hyperref}
\usepackage{color}
\usepackage{indentfirst}
\usepackage{subfigure}
\usepackage{psfrag}
\usepackage{slashed}
\usepackage{float}
\usepackage{setspace}

\begin{document}

\title{Strange form factors of nucleon with nonlocal chiral effective Lagrangian}
\author{Fangcheng He}

\affiliation{Institute of High Energy Physics, CAS, P. O. Box
918(4), Beijing 100049, China}
\affiliation{University of Chinese Academy of Sciences, Beijing 100049, China}
\author{P. Wang}

\affiliation{Institute of High Energy Physics, CAS, P. O. Box
918(4), Beijing 100049, China}
\affiliation{Theoretical Physics Center for Science Facilities,
CAS, Beijing 100049, China}

\begin{abstract}
The strange form factors of nucleon are studied with the nonlocal chiral effective Lagrangian.
One loop contributions from both octet and decuplet intermediate states are included.
The relativistic regulator is obtained by the nonlocal Lagrangian where the gauge link is introduced to guarantee the local
gauge symmetry. With the kaon loop, the calculated charge form factor is positive, while the magnetic form factor is negative. The strange magnetic moment is $-0.041^{+0.012}_{-0.014}$ with $\Lambda=0.9 \pm 0.1$ determined from the nucleon electromagnetic form factors. Our results are comparable with the recent lattice simulation.
\end{abstract}


\maketitle

\section{Introduction}
It is well known that a complete characterization of nucleon substructure must go beyond three valence
quarks. One of the great challenges of modern hadron physics is to unravel the precise role of hidden flavours in the 
structure of the nucleon. Strange quark contribution to the nucleon form factors has attracted a lot of interest because it is purely from the sea quark. The role of the sea remains a central issue in QCD, especially with respect to lattice QCD. There such terms involve so-called disconnected graphs, i.e. quark loops are connected only by gluons to the valence quarks.

Parity-violating electron scattering (PVES) has proven to be a valuable tool for experimentally determining the
strange quark contribution to the electromagnetic form factors of the proton. Under the assumption of charge
symmetry, one can deduce the strange electric or magnetic form factor $G^s_{E,M} (Q2)$ from measurements of the
corresponding proton and neutron electromagnetic form factors and the neutral-weak vector form factor of the
proton, through its contribution to PVES. While PVES measurements are very challenging, a number of groups
have succeeded, starting with SAMPLE at Bates \cite{Spayde:2003nr} and then A4 at Mainz \cite{Maas:2004ta,Maas:2004dh} and 
G0 \cite{Armstrong:2005hs}and HAPPEX \cite{Acha:2006my,Aniol:2005zg,Aniol:2004hp} at Jefferson Lab. Up to now, the experiments have 
not provided an unambiguous confirmed answer to the sign of the strange form factors, although global analyses 
do tend to suggest that $G_s^M(0) < 0$ is favoured \cite{Young:2006jc,Gonzalez-Jimenez:2014bia}.

Theoretically, though QCD is the fundamental theory to describe strong interactions, it is difficult to study hadron physics 
using QCD directly. There are many phenomenological models, such as the cloudy bag model \cite{Lu:1997sd}, the constituent 
quark model \cite{Berger:2004yi,JuliaDiaz:2003gq}, the 1/Nc expansion approach \cite{Buchmann:2002et}, the perturbative chiral quark model 
\cite{Cheedket:2002ik,Lyubovitskij:2002ng}, the extended vector meson dominance model \cite{Williams:1996id}, the SU(3) chiral quark 
model \cite{Shen:1997jd}, 
the quark-diquark model \cite{Jakob:1993th,Hellstern:1995ri}, etc. Besides the above phenomenological models, heavy baryon and 
relativistic chiral perturbation theory have been widely applied to study the hadron spectrum and structure. 
Historically, most formulations of ChPT are based on dimensional or infrared regularisation. Though ChPT is a successful and 
systematic approach, for the nucleon electromagnetic form factors, it is only valid for $Q^2 < 0.1$ GeV$^2$ \cite{Fuchs:2003ir}. 
When vector mesons are included, the result is close to the experiments with $Q^2$ less than 0.4 GeV$^2$ \cite{Kubis:2000zd}.

An alternative regularization method, namely finite-range-regularization (FRR) has been proposed. Inspired by
quark models that account for the finite-size of the nucleon as the source of the pion cloud, effective field theory
with FRR has been widely applied to extrapolate the vector meson mass, magnetic moments, magnetic form factors,
strange form factors, charge radii, first moments of GPDs, nucleon spin, etc \cite{Young:2002ib,Leinweber:2003dg,Wang:2007iw,Wang:2010hp,Allton:2005fb,
Armour:2008ke,Hall:2013oga,Leinweber:2004tc,Wang:1900ta,Wang:2012hj,Wang:2013cfp,Hall:2013dva,Wang:2015sdp,Li:2015exr,Li:2016ico,Wang:2008vb}. In the finite-range-regularization,
there is no cut for the energy integral. The regulator is not covariant and is in three-dimensional momentum space. This
non-relativistic regulator can only be applied with the heavy baryon ChPT. 

We proposed a relativistic version for the finite-range-regularization which makes it possible to study the hadron
properties with relativistic chiral effective Lagrangian at large $Q^2$ \cite{Wang:2014tna,He:2017viu}. The covariant regulator 
was generated from the nonlocal gauge invariant Lagrangian. As a result, the renomalized charge of proton (neutron) 
is 1 (0) with the additional diagrams obtained by the expansion of the gauge link. The nonlocal interaction generates 
both the regulator which makes the loop integral convergent and the $Q^2$ dependence of form factors at tree level. 
The obtained electric and magnetic form factors of nucleon are very close to the experimental data \cite{He:2017viu}. This was the first time to calculate the form factors precisely at relatively large $Q^2$ with chiral effective Lagrangian. 

In this paper, we will apply the nonlocal chiral effective Lagrangian to study the strange form factors. The paper is 
organised in the following way. In section II, we briefly introduce the chiral effective Lagrangian. The strange form factors
are derived in section III. Numerical results are presented in section IV and finally, section V is a summary.

\section{Chiral  Effective Lagrangian}
The lowest order chrial lagrangian for baryons, pseudoscalar mesons and their interaction
can be written as \cite{Jenkins:1991ts,Jenkins:1992pi}.
\begin{eqnarray}
\mathcal{L} &=& i\, Tr\,\bar{B} \gamma_{\mu}\,\slashed{\mathscr{D}}B -m_B\,Tr\,\bar{B}B
+\bar{T}_\mu^{abc}(i\gamma^{\mu\nu\alpha}D_\alpha\,-\,m_T\gamma^{\mu\nu})T_\nu^{abc}
+\frac{f^2}{4}Tr\,\partial_\mu\Sigma\partial^\mu\Sigma^+ +D\,Tr\, \bar{B} \gamma_\mu \gamma_5\, \{A_\mu,B\} \nonumber \\ 
 &+& F\,Tr\, \bar{B} \gamma_\mu \gamma_5\, [A_\mu,B]
 +\left [\frac{{\cal C}}{f}\epsilon^{abc}\bar{T}_\mu^{ade}(g^{\mu\nu}+z\gamma_\mu\gamma_\nu) B_c^e\partial_\nu\phi_b^d
 +H.C \right ]
 +H\bar{T}^\mu\gamma_\nu\gamma^5A^\nu\,T_\mu,
\end{eqnarray}
where $D$, $F$, $\cal C$ and $H$ are the coupling constants.
The chiral covariant derivative $\mathscr{D}_\mu$ is defined as $\mathscr{D}_\mu
B=\partial_\mu B+[V_\mu,B]$. The pseudoscalar meson octet
couples to the baryon field through the vector and axial vector
combinations as
\begin{equation}
V_\mu=\frac12(\zeta\partial_\mu\zeta^\dag+\zeta^\dag\partial_\mu\zeta),~~~~
A_\mu=\frac12(\zeta\partial_\mu\zeta^\dag-\zeta^\dag\partial_\mu\zeta),
\end{equation}
where
\begin{equation}
\Sigma=\zeta^2=e^{2i\phi/f}, ~~~~~~ f=93~{\rm MeV}.
\end{equation}
The matrix of pseudoscalar fields $\phi$ is expressed as
\begin{eqnarray}
\phi=\frac1{\sqrt{2}}\left(
\begin{array}{lcr}
\frac1{\sqrt{2}}\pi^0+\frac1{\sqrt{6}}\eta & \pi^+ & K^+ \\
~~\pi^- & -\frac1{\sqrt{2}}\pi^0+\frac1{\sqrt{6}}\eta & K^0 \\
~~K^- & \bar{K}^0 & -\frac2{\sqrt{6}}\eta
\end{array}
\right).
\end{eqnarray}
$\mathscr{A}^\mu$ is the photon field. 
The covariant derivative $D_\mu$ in the decuplet part is defined as
$D_\nu T_\mu^{abc} = \partial_\nu T_\mu^{abc}+(\Gamma_\nu,T_\mu)^{abc}$,  where $\Gamma_\nu$ 
is the chrial connection\cite{Scherer:2002tk} defined as $(X,T_\mu)=(X)_d^aT_\mu^{dbc}+(X)_d^bT_\mu^{adc}+(X)_d^c
T_{\mu}^{abd}$. 
$\gamma^{\mu\nu\alpha}$,$\gamma^{\mu\nu}$ are the antisymmetric matrices expressed as
\begin{equation}
\gamma^{\mu\nu}
=\frac12\left[\gamma^\mu,\gamma^\nu\right]\hspace{.5cm}\text{and}\hspace{.5cm}
\gamma^{\mu\nu\rho}=\frac14\left\{\left[\gamma^\mu,\gamma^\nu\right],
\gamma^\rho\right\}\,
\end{equation}
In the chiral $SU(3)$ limit, the octet and decuplet baryons will have the same
mass $m_B$ and $m_T$. In our calculation, we use the physical masses for
baryon octets and decuplets. The explicit form of the baryon octet
is written as
\begin{eqnarray}
B=\left(
\begin{array}{lcr}
\frac1{\sqrt{2}}\Sigma^0 +\frac1{\sqrt{6}}\Lambda &
\Sigma^+ & p \\
~~\Sigma^- & -\frac1{\sqrt{2}}\Sigma^0+\frac1{\sqrt{6}}\Lambda & n \\
~~\Xi^- & \Xi^0 & -\frac2{\sqrt{6}}\Lambda
\end{array}
\right).
\end{eqnarray}
For the baryon decuplets, there are three indices, defined as
\begin{eqnarray}
T_{111}=\Delta^{++}, ~~ T_{112}=\frac1{\sqrt{3}}\Delta^+, ~~
T_{122}=\frac1{\sqrt{3}}\Delta^0,\nonumber\\T_{222}=\Delta^-, ~~
T_{113}=\frac1{\sqrt{3}}\Sigma^{\ast,+}, ~~
T_{123}=\frac1{\sqrt{6}}\Sigma^{\ast,0}, \nonumber\\
T_{223}=\frac1{\sqrt{3}}\Sigma^{\ast,-}, ~~
T_{133}=\frac1{\sqrt{3}}\Xi^{\ast,0}, ~~
T_{233}=\frac1{\sqrt{3}}\Xi^{\ast,-}, ~~ T_{333}=\Omega^{-}.
\end{eqnarray}
The octet, decuplet and octet-decuplet transition magnetic moment
operators are needed in the one loop calculation of nucleon electromagnetic
form factors. The baryon octet anomalous magnetic Lagrangian is written as
\begin{equation}\label{lomag}
{\cal L}=\frac{e}{4m_N}\left(c_1{\rm Tr}\bar{B} \sigma^{\mu\nu}
\left\{F^+_{\mu\nu},B\right\}+c_2{\rm Tr}\bar{B}
\sigma^{\mu\nu} \left[F^+_{\mu\nu},B \right]\right),
\end{equation}
where
\begin{equation}
F^+_{\mu\nu}=-\frac12\left(\zeta^\dag F_{\mu\nu}Q\zeta+\zeta
F_{\mu\nu}Q\zeta^\dag\right).
\end{equation}
At the lowest order, the Lagrangian will generate the following nucleon anomalous magnetic moments:
\begin{equation}\label{treemag}
F_2^p=\frac13 c_1+c_2,~~~~~~ F_2^n=-\frac23 c_1.
\end{equation}
The transition magnetic operator is
\begin{equation}
{\cal L}=i\frac{e}{4m_N}\mu_TF_{\mu\nu}(\epsilon_{ijk}Q^i_j\bar{B}^j_m\gamma^\mu\gamma_5T^{\nu,klm}
+\epsilon^{ijk}Q^l_i\bar{T}^\mu_{klm}\gamma^\nu\gamma_5B^m_j),
\end{equation}
where the matrix $Q$ is defined as $Q=$diag$\{2/3,-1/3,-1/3\}$. 
The effective decuplet anomalous magnetic moment operator can be  expressed as effective Lagrangian
\begin{eqnarray}\label{eq:ci}
{\cal L}=\frac{ieF_2^T}{2m_T}\bar{T}_\mu^{abc}\sigma^{\rho\sigma}q_{\sigma}\mathscr{A}_\rho T_\mu^{abc}.
\end{eqnarray}
In quark model, the baryon magnetic moments can also be written in terms of quark magnetic moments. For example,
$\mu_p=\frac43 \mu_u-\frac13\mu_d$, $\mu_n=\frac43 \mu_d-\frac13\mu_u$, $\mu_{\Delta^{++}}\,=\,3\mu_u$. 
Using $\mu_u=-2\mu_d=-2\mu_s$, $\mu_T$, $F_2^T$ and $\mu_q$ can be written in terms of $c_1$. For example,
\begin{eqnarray}\label{eq:c1}
\mu_T=4c_1, ~~~~~ F_2^{\Delta^{++}}\,=2c_1-2, ~~~~~ \mu_s = -\frac13 c_1.
\end{eqnarray}
The strange quark contribution to the hyperons at tree level can be written as \cite{Ha:2002sa}
\begin{eqnarray}
\mu_{\Sigma^+}^s = \mu_{\Sigma^-}^s = \mu_{\Sigma^0}^s = \mu_s,~~~~~
\mu_\Lambda^s = -3\mu_s.
\end{eqnarray}
Similarly, the strange quark contribution to the decuplet and transition magnetic moments at tree level can be 
written as \cite{Ha:1998gg}
\begin{eqnarray}
\mu_{\Sigma^{*,+}}^s &=& \mu_{\Sigma^{*,-}}^s = \mu_{\Sigma^{*,0}}^s = -3\mu_s,    \\
\mu_{\Sigma^{*,+}\Sigma^+}^s &=& -\mu_{\Sigma^{*,-}\Sigma^-}^s = -\mu_{\Sigma^{*,0}\Sigma^0}^s = -2\sqrt{2}\mu_s,
\end{eqnarray}
Following the usual convention, the charge of the strange quark is taken to be $1$.

Now we construct the nonlocal Lagrangian which will generate the covariant regulator. 
The gauge invariant non-local Lagrangian can be obtained using the method in \cite{Terning:1991yt,Wang:2014tna,He:2017viu}. 
For instance, the local interaction including kaon can be written as 
\begin{equation}
{{\cal L}_K}^{local}=-\int\!\,dx \frac{D+3F}{\sqrt{12}f} \bar{p}(x)\gamma^\mu\gamma_5\,\Lambda(x)(\partial_\mu+ie\,
\mathscr{A}_\mu^s(x)) K^+(x),
\end{equation}
where $\mathscr{A}_\mu^s(x)$ is the external field interacting the strange quark.
The nonlocal Lagrangian for this interaction is expressed as 
\begin{align}\label{eq:nonlocal}
{{\cal L}_K}^{nl}&=-\int\!\,dx\int\!\,dy\frac{D+F}{\sqrt{12}f}\bar{p}(x)\gamma^\mu\gamma_5 \Lambda(x)(\partial_\mu+i\,e\int\!\,
da\mathscr{A}_\mu^s(x-a)F(a))                  \nonumber\\
&\times\text {exp}[ie\int_x^y dz^\nu\,\int\!\,da\mathscr{A}_\nu^s(z-a)F(a)]\,K^+(y)F(x-y),
\end{align}
where $F(x)$ is the correlation function.
To guarantee the gauge invariant, the gauge link is introduced in the above Lagrangian.
The regulator can be generated automatically with correlation function.
With the same idea, the nonlocal interaction between baryons and $\mathscr{A}_\mu^s(x)$ can also be obtained. 
For example, the local interaction between $\Lambda$ and external field is written as
\begin{align}
{\cal L}_{\Lambda}^{local} = & -e \bar{\Lambda}(x) \gamma^\mu \Lambda(x) \mathscr{A}_\mu^s(x)
+\frac{(c_1-1)e}{4m_\Lambda} \bar{\Lambda}(x)\sigma^{\mu\nu}\Lambda(x)F_{\mu\nu}^s(x).
\end{align}
The corresponding nonlocal Lagrangian is expressed as
\begin{align}
 {\cal L}_{\Lambda}^{nl} = -e \int da \bar{\Lambda}(x) \gamma^\mu \Lambda(x) \mathscr{A}_\mu^s(x-a)F_1(a)
+ \frac{(c_1-1)e}{4m_\Lambda}\int da \bar{\Lambda}(x)\sigma^{\mu\nu}\Lambda(x)F_{\mu\nu}^s(x-a)F_2(a),
\end{align} 
where $F_1(a)$ and $F_2(a)$ is the correlation function for the nonlocal strange electric and magnetic interactions.
The form factors at tree level which are momentum dependent can be easily obtained with the Fourier transformation
of the correlation function.
The simplest choice is to assume that the correlation function of the strange electromagnetic vertex is the same 
as that of the lambda-kaon vertex, i.e. $F_1(a)=F_2(a)=F(a)$. 
Therefore, the Dirac and Pauli form factors will have the same dependence on the momentum transfer at tree level.
The better choice is to assume that the charge and magnetic form factors
at tree level have the same the momentum dependence as lambda-kaon vertex, 
i.e. $G_M^{\rm tree}(p)=c_1G_E^{\rm tree}(p) = c_1\tilde{F}(p)$,
where $\tilde{F}(p)$ is the Fourier transformation of the correlation function $F(a)$ \cite{He:2017viu}.
The corresponding function of $\tilde{F}_1(q)$ and $\tilde{F}_2(q)$ is then expressed as
\begin{eqnarray}
\tilde{F}_1^p(q)&=&\tilde F(q)\frac{4m_\Lambda^2+c_1Q^2}{4m_\Lambda^2+Q^2},~~~\tilde{F}_2^p(q)=\tilde F(q)
\frac{4m_\Lambda^2}{4m_\Lambda^2+Q^2}, 
\end{eqnarray}
From the above equations, one can see that in the heavy baryon limit, these two choices are equivalent.
The nonlocal Lagrangian is invariant under the following gauge transformation 
\begin{eqnarray}
K^+(y)\rightarrow e^{-i\alpha(y)}K^+(y),~~~~\Lambda(x)\rightarrow e^{i\alpha(x)}\Lambda(x), ~~~~\mathscr{A}_\mu^s(x)
\rightarrow\,\mathscr{A}_\mu^s(x)+\frac1e\partial_\mu\alpha^\prime(x),
\end{eqnarray}
where $\alpha(x)=\int\!\,da\,\alpha^\prime(x-a)F(a)$. From Eq.~(\ref{eq:nonlocal}), two kinds of couplings
between hadrons and external field can be obtained. One is the normal one expressed as
\begin{equation}
{\cal L}^{nor}=-ie\int\!\,dx\int\!\,dy\frac{D+F}{\sqrt{12}f}\bar{p}(x)\gamma^\mu\gamma_5 \Lambda(x)F(x-y)K^+(y)
\int\!\,da\mathscr{A}_\mu^s(x-a)F(a),
\end{equation}
This interaction is similar as the traditional local Lagrangian except the correlation function. 
The other one is the additional interaction obtained by the expansion of the gauge link, expressed as
\begin{equation}
{\cal L}^{add}=-ie\int\!\,dx\int\!\,dy\frac{D+F}{\sqrt{12}f}\bar{p}(x)\gamma^\mu\gamma_5 \Lambda(x)\partial_\mu
\left(F(x-y)\int_x^y dz^\nu\int\!\,da\mathscr{A}_\nu^s(z)F(z-a)\,K^+(y)\right)                     
\end{equation}
The additional interaction is important guarantee the local gauge symmetry resulting the net strangeness of nucleon zero.

\section{STRANGE FORM FACTORS}

The strange quark contribution to the Dirac and Pauli form factors are defined as
\begin{equation}\label{eq:f1f2}
<N(p')|J_\mu^s|N(p)>=\bar{u}(p')\left\{\gamma^\mu
F_1^s(Q^2)+\frac{i\sigma^{\mu\nu}
q_\nu}{2m_N}F_2^s(Q^2)\right\}u(p),
\end{equation}
where $q=p^\prime-p$ and $Q^2=-q^2$.
The combination of the above form factors can generate electric and magnetic form factors contributed from strange quark
\begin{equation}
G_E^s(Q^2) =F_1^s(Q^2)-\frac{Q^2}{4m_N^2}F_2^s(Q^2), ~~~~ G_M^s(Q^2) =F_1^s(Q^2) +F_2^s(Q^2).
\end{equation}
According to the Lagrangian, the one loop Feynman diagrams which contribute to the strange form factors are plotted in Fig~1. 
\begin{figure}[tbp]
\begin{center}
\includegraphics[scale=0.85]{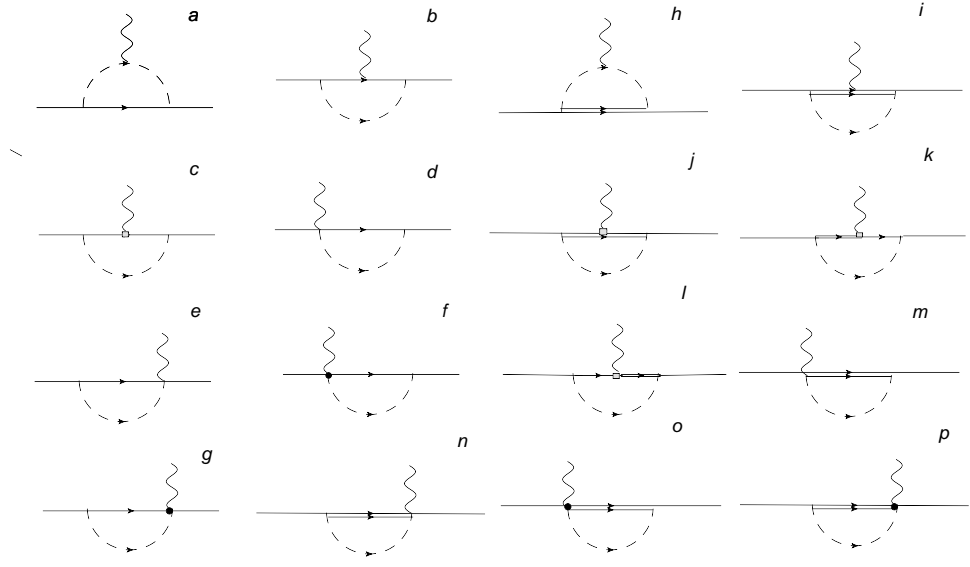}
\caption{One-loop contributions to the proton strange electromagnetic form factor. The solid, double-solid, dashed and wave lines 
are for the octet baryons, decuplet baryons, pseudoscalar mesons and photons, respectively. The rectangle and 
blackdot respresent magentic and additional interacting vertex.}
\label{diagrams}
\end{center}
\end{figure}

In this section, we will only show the expressions for the intermediate octet baryon part. For the intermediate decuplet baryon part, 
the expressions are written in the Appendix. In Fig.~1a, the external field couples to the meson. The contribution 
of Fig.~1a to the matrix element in Eq.~(\ref{eq:f1f2}) is expressed as
\begin{eqnarray} 
\Gamma_a^{\mu(p)}&=&\frac{(3F+D)^2}{12f^2}I_{aK}^{\Lambda}+\frac{3(D-F)^2}{4f^2}I_{aK}^{\Sigma},
\end{eqnarray} 
where the integral $I_{aK}^\Lambda$ is expressed as
\begin{equation} 
I_{aK}^\Lambda=\bar{u}(p')\tilde F(q)\int\!\frac{d^4 k}{(2\pi)^4}(\slashed{k}+\slashed{q})\gamma_5\,\tilde F(q+k)
\frac{1}{D_K(k+q)}\,(2k+q)^\mu\frac{1}{D_K(k)}\frac{1}{\slashed{p}-\slashed{k}-m_\Lambda}
(-\slashed{k}\gamma_5)\tilde F(k)u(p)
\end{equation}
and $D_K(k)$ is defined as 
\begin{equation} 
D_K(k)=k^2-M_k^2+i\epsilon.
\end{equation}
$m_\Lambda$ and $M_k$ are the masses for the intermediate $\Lambda$ hyperon and $K$ meson, respectively. 
The integral $I_{aK}^\Sigma$ is the same as $I_{aK}^\Lambda$ except the intemediate hyperon mass $m_\Lambda$
is replaced by $m_\Sigma$. Therefore, here we only show the expressions for $\Lambda$ hyperon.
In Fig.1b, the external field couples to the intermediate hyperons with electric interaction. 
The contribution of this diagram is expressed as 
\begin{eqnarray} 
\Gamma_b^{\mu(p)}&=&\frac{(3F+D)^2}{12f^2}\frac{4m_\Lambda^2+c_1Q^2}{4m_\Lambda^2+Q^2}I_{bK}^{\Lambda}
+\frac{3(D-F)^2}{4f^2}\frac{12m_\Sigma^2-c_1Q^2}{12m_\Sigma^2+3Q^2}I_{bK}^{\Sigma},
\end{eqnarray} 
where the integral $I_{bK}^{\Lambda}$ is written as
\begin{equation} 
I_{bK}^{\Lambda}=\bar{u}(p')\tilde F(q)\int\!\frac{d^4 k}{(2\pi)^4}\slashed{k}\gamma_5\,\tilde F(k)\frac{1}{D_K(k)}
\frac{1}{\slashed{p'}-\slashed{k}-m_\Lambda}\gamma^\mu\frac{1}{\slashed{p}-\slashed{k}
-m_\Lambda}\frac{\slashed{k}\gamma_5}{\sqrt{2}f}\tilde F(k)u(p).
\end{equation}
Fig.1c is similar as Fig.1b except for the magnetic interaction. The contribution of this diagram is written as
\begin{eqnarray} 
\Gamma_c^{\mu(p)}&=&\frac{(3F+D)^2}{12f^2}\frac{4(c_1-1)m_\Lambda^2}{4m_\Lambda^2+Q^2}
I_{cK}^{\Lambda}-\frac{3(D-F)^2}{4f^2}\frac{(4c_1+12)m_\Sigma^2}{12m_p^2+3Q^2}I_{cK}^{\Sigma},                         
\end{eqnarray} 
where $I_{cK}^\Lambda$ is expressed as
\begin{eqnarray}
I_{cK}^{\Lambda}&=&\bar{u}(p')\tilde F(q)\int\!\frac{d^4 k}{(2\pi)^4}\slashed{k}\gamma_5\tilde F(k)
\frac{1}{\slashed{p'}-\slashed{k}-m_\Lambda}
\frac{\sigma^{\mu\nu}q_{\nu}}{2m_\Lambda}\frac{1}{\slashed{p}-\slashed{k}-m_\Lambda}\frac{i}{D_K(k)}
\slashed{k}\gamma_5\tilde F(k)u(p).
\end{eqnarray}
Fig.~1d and 1e are the Kroll-Ruderman diagrams. The contribution from these two diagrams is written as 
\begin{eqnarray} 
\Gamma_{d+e}^{\mu(p)}&=&\frac{(3F+D)^2}{12f^2}I_{(d+e)K}^{\Lambda}+\frac{3(D-F)^2}{4f^2}I_{(d+e)K}^{\Sigma},
\end{eqnarray} 
where
\begin{eqnarray}
I_{(d+e)K}^{\Lambda}&=&\bar{u}(p')\tilde F(q)\int\!\frac{d^4 k}{(2\pi)^4}\slashed{k}\gamma_5\tilde F(k)
\frac{1}{\slashed{p'}-\slashed{k}-m}\frac{1}{D_K(k)}\gamma^\mu\gamma_5 \tilde F(k)u(p)     \nonumber\\
&+&\bar{u}(p')\tilde F(q)\int\!\frac{d^4 k}{(2\pi)^4}\gamma^\mu\gamma_5\tilde F(k)\frac{1}{\slashed{p}-\slashed{k}-m}
\frac{1}{D_K(k)}\slashed{k}\gamma_5\tilde F(k)u(p).
\end{eqnarray} 
These two diagrams only have contribution in the relativistic cases. In the heavy baryon limit, 
they have no contribution to either electric or magnetic form factors.
Fig.~1f and 1g are the additional diagrams which generated from the expansion of the gauge link.
The contribution of these two additional diagrams are expressed as
\begin{eqnarray} 
\Gamma_{f+g}^{\mu(p)}&=&\frac{(3F+D)^2}{12f^2}I_{(f+g)K}^{\Lambda}+\frac{3(D-F)^2}{4f^2}I_{(f+g)K}^{\Sigma},
\end{eqnarray} 
where
\begin{eqnarray}
I_{(f+g)K}^{\Lambda}&=&\,\bar{u}(p')\tilde F(q)\int\!\frac{d^4 k}{(2\pi)^4}\slashed{k}\gamma_5\tilde F(k)\frac{1}{\slashed{p'}
-\slashed{k}-m_\Lambda}\frac{1}{D_K(k)}(-\slashed{k}+\slashed{q})\gamma_5\frac{(2k-q)^\mu}{2kq-q^2}
[\tilde F(k-q)-\tilde F(k)]u(p) \nonumber\\
&+&\bar{u}(p')\tilde F(q)\int\!\frac{d^4 k}{(2\pi)^4}(\slashed{k}+\slashed{q})\gamma_5\frac{(2k+q)^\mu}{2kq+q^2}
[\tilde F(k+q)-\tilde F(k)]
\frac{1}{\slashed{p}-\slashed{k}-m_\Lambda}\frac{1}{D_K(k)}\slashed{k}\gamma_5\tilde F(k)u(p).
\end{eqnarray} 
Using FeynCalc to simplify the $\gamma$ matrix algebra, we can get the separate expressions for the Dirac and Pauli form factors.

\begin{figure}[t]
\begin{center}
\includegraphics[scale=0.75]{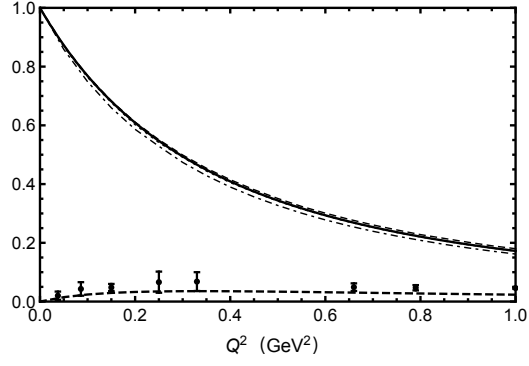}
\caption{Nucleon electromagnetic form factors versus momentum transfer $Q^2$. The solid line is for the 
empirical function $1/(1 +Q^2/0.71$GeV$^2)^2$. The dashed, dotted and dash-dotted
lines are for $G_E^p$, $G_M^p/\mu_p$ and $G_M^n/\mu_n$, respectively. The dotted line and solid line coincide.
The dashed line started from 0 is for $G_E^n$.
The experimental data of neutron charge form factor are from Ref.~\cite{Seimetz:2005vg}.}.
\label{diagrams}
\end{center}
\end{figure}

\section{Numerical Results}
In the numerical calculations, the parameters are chosen as $D=0.76$
and $F=0.5$ ($g_A=D+F=1.26$). The coupling constant ${\cal C}$ is
chosen to be $1$ which are the same as \cite{Pascalutsa:2006up}.
The off-shell parameter $z$ is chosen to be $z=-1$ \cite{Nath:1971wp}. 
The covariant regulator is chosen to be of a dipole form 
\begin{equation}
\tilde F(k)=\frac{\Lambda^4}{(k^2-M_j^2-\Lambda^2)^2},
\end{equation} 
where $M_j$ is the mass of the corresponding meson and it is zero for photon.
Therefore, in this nonlocal Lagrangian, there are three parameters $c_1$, $c_2$ and $\Lambda$ to be determined. 
$\Lambda$ is chosen to get the best description of the nucleon form factors up to relatively large $Q^2$.
By comparing with the experimental electromagnetic form factors of nucleon, the best $\Lambda$ is found to be
around 0.9 GeV.  The other two parameters $c_1$ and $c_2$ are determined by the experimental magnetic moments 
of proton and neutron. With $\mu_p= 2.79$ and $\mu_n=-1.91$, we get $c_1=2.081$ and $c_2=0.788$.

Before present the results for strange form factors, we first show the electromagnetic form factors.
In Fig.~2, the charge and normalized magnetic form factors of proton and neutron with $\Lambda=0.9$ GeV are plotted.
The solid line is for the empirical function $1/(1 +Q^2/0.71 $GeV$^2)^2$. The dashed, dotted and dash-dotted
lines are for $G_E^p$, $G_M^p/\mu_p$ and $G_M^n/\mu_n$, respectively. The dotted line is invisible because it coincides with the empirical line. The dashed line started from 0 is for $G_E^n$.
The experimental data of neutron charge form factor are from Ref.~\cite{Seimetz:2005vg}. From the figure, we can see that our calculated form
factors are very close to the experimental data which is a great improvement compared with the results
of the traditional chiral effective field theory \cite{Fuchs:2003ir,Kubis:2000zd}.

\begin{figure}[t]
\begin{center}
\includegraphics[scale=0.85]{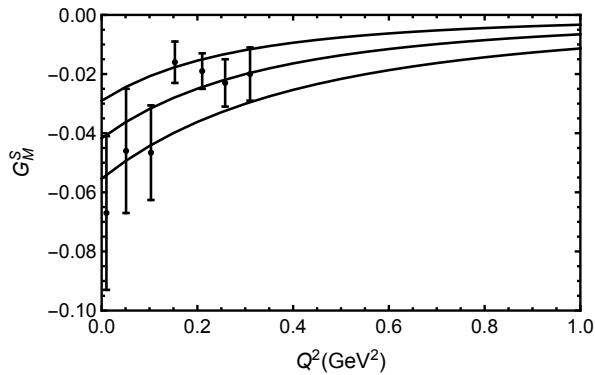}
\caption{The strange magnetic form factor of nucleon versus momentum transfer $Q^2$ with different $\Lambda$. 
The three solid lines from bottom to top, are for the results with $\Lambda=$1 GeV, 0.9 GeV, 0.8 GeV, respectively. 
The data with error bars are from Lattice simulation \cite{Sufian:2017osl}.}
\label{diagrams}
\end{center}
\end{figure}

\begin{figure}[t]
\begin{center}
\includegraphics[scale=0.85]{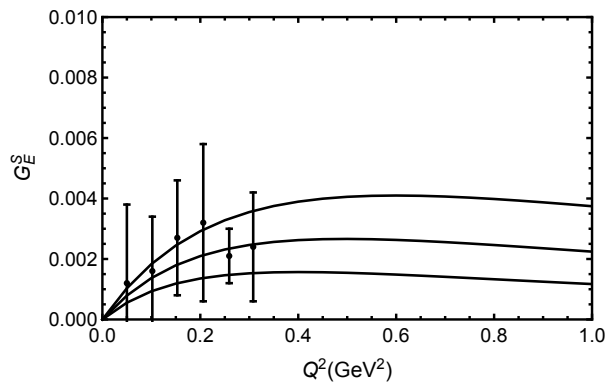}
\caption{The strange electric form factor of nucleon versus momentum transfer $Q^2$ with different $\Lambda$. 
The three solid lines from top to bottom, are for the results with $\Lambda=$1 GeV, 0.9 GeV, 0.8 GeV, respectively. 
The data with error bars are from Lattice simulation \cite{Sufian:2017osl}.}
\label{diagrams}
\end{center}
\end{figure}

Now we show the results for the strange form factors. 
The strange magnetic form factor $G_M^s(Q^2)$ of nucleon versus $Q^2$ with different $\Lambda$ is plotted in Fig.~3.
The three solid lines from bottom to top, are for the results with $\Lambda=$1 GeV, 0.9 GeV and 0.8 GeV, respectively. 
The data with error bars from recent lattice simulation \cite{Sufian:2017osl} are also shown in the figure. The strange magnetic form factors
increases with the increasing momentum transfer $Q^2$. At zero momentum transfer,
when $\Lambda=0.9\pm 0.1$ GeV, $G_M^s(0)=-0.041_{-0.014}^{+0.012}$. The absolute value of strange magnetic moment in this relativistic chiral Lagrangian is smaller than that in heavy baryon approach, where 
$G_M^s(0)=-0.058_{-0.053}^{+0.034}$ \cite{Wang:2013cfp}. The main reason for the difference is that rainbow diagrams (Fig.~1a and Fig.~1c) have much smaller contribution to $G_M^s$ than that in the heavy baryon limit due to the covariant regulator. 
Though the Kroll-Ruderman and additional diagrams have sizeable contribution to $G_M^s$ in this relativistic case, while in the heavy baryon limit such contribution is zero, the total absolute value of $G_M^s(0)$ is a little smaller in relativistic case.

The strange charge form factor $G_E^s(Q^2)$ is plotted in Fig.~4. The three solid lines from top to bottom, are for the results 
with $\Lambda=1$ GeV, 0.9 GeV and 0.8 GeV, respectively. When $Q^2=0$, $G_E^s(0)=0$. This is true only when
the additional diagrams generated from the expansion of the gauge link are included. The strange charge form factor first
increases and then decreases with the increasing $Q^2$. At finite $Q^2$, $G_E^s(Q^2)$ is always a small positive number.
It is clear that for both strange charge and magnetic form factors, our result is comparable with the Lattice data. 
With the strange form factors, the strange radii can be obtained as
\begin{equation}
<(r_E^s)^2>=-6{dG_E^s(Q^2)\over{dQ^2}}|_{Q^2=0},\,\,\,\,\,\,\,\,\,\,\,\,\,\, 
<(r_M^s)^2>=-6{dG_M^s(Q^2)\over{dQ^2}}|_{Q^2=0}.
\end{equation}
With $\Lambda =0.9\pm 0.1$ GeV, we have $<(r_E^s)^2> = -0.004\pm 0.001$ fm$^2$ and $<(r_M^s)^2> = -0.028\pm 0.003$ fm$^2$.

\begin{figure}
\begin{center}
\includegraphics[scale=0.85]{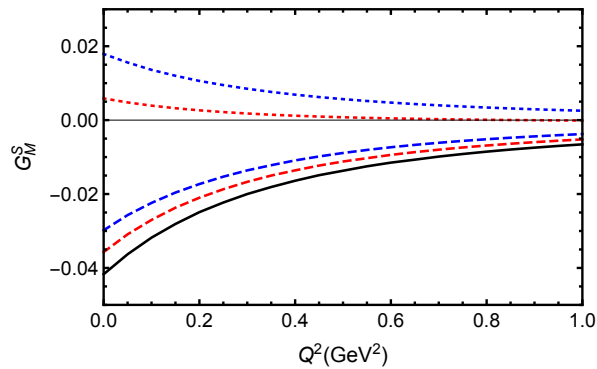}
\caption{The proton strange magnetic form factor versus momentum transfer $Q^2$ with  $\Lambda=0.9$ GeV. The solid, dashed 
and dotted lines are for total, octet and decuplet contribution to $G_M^s(Q^2)$, respectively. The red lines are for the normal 
diagrams and the blue lines are for the additional diagrams.}
\label{diagrams}
\end{center}
\end{figure}

\begin{figure}
\begin{center}
\includegraphics[scale=0.87]{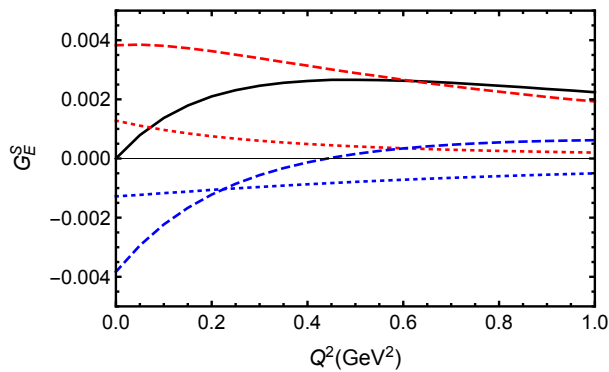}
\caption{Same as Fig.~5, but for strange charge form factor $G_M^s(Q^2)$.}
\label{diagrams}
\end{center}
\end{figure}

To see clearly the separate contribution from the octet and decuplet parts, and from the normal diagrams and additional 
diagrams, in Fig.~5, we plot each contribution to the strange magnetic form 
factor at $\Lambda=0.9$ GeV separately. The solid, dashed and dotted lines are for total, octet and decuplet contribution to
$G_M^s(Q^2)$, respectively. The red lines are for the contribution from normal diagrams 
and the blue lines are for the contribution from additional diagrams.
From the figure, one can see that, the octet contribution is dominant. Compared with the octet contribution, the decuplet part 
gives a smaller opposite number to $G_M^s$. The additional diagrams also provide important contributions to the total
$G_M^s$. 

In Fig.~6, we plot the same curves, but for the strange charge form factor. At $Q^2=0$, the contributions from the normal
and additional diagrams cancel each other. As a result, the net strangeness is zero. This is guaranteed by the gauge symmetry
of strange quark. Similar as in the magnetic case, the octet contribution is dominant for the total $G_E^s(Q^2)$.
At small $Q^2$, the contribution from the additional diagrams changes more quickly than that from the normal diagrams. Therefore, $G_E^s$ first increases from 0 and then decreases smoothly with the increasing $Q^2$.

\begin{table}
\begin{center}
\caption{Contributions to the strange magnetic moment from each diagram and total $G_M^s(0)$ in unit of $\mu_N$.
 The results from the heavy baryon formalism \cite{Wang:2013cfp} are also listed in the last line.}
\begin{ruledtabular}
\begin{tabular}{ccccccccccccc}
 $\Lambda$ (GeV) &$1a$&$1b$& $1c$& $1d+1e$ &$1f+1g$ &$1h$& $1i$ &$1j$&$1k+1l$& $1m+1n$&$1o+1p$
 &$G_M^s(0)$                                                                                      \\  \hline
0.8 &$-0.011$ &0.003 &  $-0.005$ & $-0.010$ & $-0.021$ & 0.005 &$-0.0002$ & 0.005 & $-0.007$ & 0.0007& 0.011 
&$-0.029$\\
0.9 & $-0.017$ &0.005 &  $-0.008$ & $-0.016$ & $-0.030$ & 0.009 & $-0.0004$ & 0.009 & $-0.012$ & 0.001 &0.018 &$-0.041$\\
1 &$-0.024$ &  0.009 & $-0.011$ & $-0.026$ & $-0.039$ &0.013 & $-0.0008$ & 0.014 & $-0.018$ & 0.002 & 0.026 &$-0.055$   
\\
\hline
0.8 & $-0.050$ & $-$ &  $-0.021$ & $-$ & $-$ & 0.009 & $-$ & 0.013 & $-0.009$ & $-$ & $-$ &$-0.058$\\
\end{tabular}
\end{ruledtabular}
\end{center}
\end{table}

\section{Summary}
We studied the strange form factors of nucleon with the nonlocal chiral effective Lagrangian. Both the octet and decuplet
intermediate states are included in the one loop calculation. The covariant form factors are derived from the nonlocal Lagrangian. This is the relativistic version of the finite-range-regularization, which make it possible to study the hadron structure at relatively large $Q^2$. From the previous study of the nucleon electromagnetic form factors, it shows this nonlocal Lagrangian method is a great improvement compared with the traditional chiral effective theory.
The gauge link is introduced to guarantee the local gauge symmetry. As a result, in addition to the normal diagrams which are generated from the minimal substitution, the additional diagrams appear which are generated from the expansion 
of the gauge link. These additional diagrams are crucial to get the net strangeness zero at $Q^2=0$ for nucleon. 
They also have important contribution to the magnetic form factors. For both $G_E^s$ and $G_M^s$, 
the octet intermediated states provide more important contribution than decuplet intermediate states.
In this nonlocal chiral effective Lagrangian, there are three free parameters. $c_1$ and $c_2$ are determined
by the experimental magnetic moments of proton and neutron. $\Lambda$ in the correlation function is determined 
by the best description of the nucleon electromagnetic form factors up to relatively large $Q^2$. At finite momentum transfer, the strange charge form
factor is positive, while the strange magnetic form factor is negative. At $Q^2=0$, the strange magnetic moment is
$-0.041_{-0.014}^{+0.012}$. Compared with the calculated $G_M^s(0)$ in heavy baryon formalism with finite-range-regularization, the absolute value of $G_M^s(0)$ calculated in this relativistic version is a little smaller.  
Our results are also comparable with the recent lattice simulation. As a summary, we list the contribution to the strange magnetic moment of each diagram in Table I.

\section*{Acknowledgments}
This work is supported by the National Natural Sciences Foundations of China under the grant No. 11475186, the 
Sino-German CRC 110 ``Symmetries and the Emergence of Structure in QCD'' project by NSFC under the grant 
No.11621131001, and the Key Research Program of Frontier Sciences, CAS under grant No. Y7292610K1.
\section*{Appendix}

The expressions for the decuplet part are written in the following way.
The contribution of Fig.~1h is expressed as
\begin{eqnarray} 
\Gamma_h^{\mu(p)}&=&\frac{{\cal C}^2}{4f^2}I_{hK}^{N\Sigma^*},
\end{eqnarray} 
where
\begin{eqnarray} 
I_{hK}^{N\Sigma^*}&=&\bar{u}(p')\tilde F(q)\int\!\frac{d^4 k}{(2\pi)^4}\left((k+q)_\sigma+z(\slashed{k}+\slashed{q})
\gamma_\sigma)\right)\tilde F(q+k)\frac{1}{D_K(k+q)}\,(2k+q)^\mu          \nonumber\\
&\times&\frac{1}{D_K(k)}\frac{1}{\slashed{p}-\slashed{k}-m_{\Sigma^*}}S_{\sigma\rho}(p-k)
\left(-k_\rho-z\gamma_\rho\slashed{k})\right)\tilde F(k)u(p).
\end{eqnarray} 
$S_{\sigma\rho}$ is expressed as
\begin{equation} 
S_{\sigma\rho}(k)=-g_{\sigma\rho}+\frac{\gamma_\sigma\gamma_\rho}{3}+\frac{2k_\sigma\,k_\rho}
{3m_{\Sigma^*}^2}+\frac{\gamma_\sigma\,k_\rho-\gamma_\rho\,k_\sigma}{3m_{\Sigma^*}}
\end{equation}
The contribution of Fig.~1i is expressed as  
\begin{equation} 
\Gamma_i^{\mu(p)}=\frac{{\cal C}^2}{4f^2}\frac{4m_{\Sigma^*}^2+c_1Q^2}{4m_{\Sigma^*}^2+Q^2}I_{iK}^{N\Sigma^*},
\end{equation}
where
\begin{eqnarray} 
I_{iK}^{N\Sigma^*}&=&\bar{u}(p')\tilde F(q)\int\!\frac{d^4 k}{(2\pi)^4}(k_\sigma+z\slashed{k}\gamma_\sigma)
\tilde F(k)\frac{1}{D_K(k)}
\frac{1}{\slashed{p'}-\slashed{k}-m_{\Sigma^*}}S_{\sigma\alpha}(p^\prime-k)    \nonumber\\
&\times&(\gamma^{\alpha\beta\mu})\frac{1}{\slashed{p}-\slashed{k}-m_{\Sigma^*}}S_{\beta\rho}
(k_\rho+z\gamma_\rho\slashed{k})F(k)u(p).
\end{eqnarray} 
The contribution of Fig.~1j is expressed as
\begin{eqnarray} 
\Gamma_j^{\mu(p)}&=&\frac{{\cal C}^2}{4f^2}\frac{4m_{\Sigma^*}^2}{4m_{\Sigma^*}^2+Q^2}I_{jK}^{N\Sigma^*},
\end{eqnarray} 
where
\begin{eqnarray} 
I_{jK}^{N\Sigma^*}&=&\bar{u}(p')\tilde F(q)\int\!\frac{d^4 k}{(2\pi)^4}(k_\sigma+z\slashed{k}\gamma_\sigma)\tilde F(k)
\frac{1}{D_K(k)}\frac{1}{\slashed{p'}-\slashed{k}-m_{\Sigma^*}}S_{\sigma\nu}(p^\prime-k)\frac{(1-c_1)}{2m_{\Sigma^*}}
\sigma^{\mu\lambda}q_\lambda \nonumber\\
 &\times&\frac{i}{\slashed{p}-\slashed{k}-m_{\Sigma^*}}S_{\nu\rho}(p-k)(k_\rho+z\gamma_\rho\slashed{k})\tilde F(k)u(p).
\end{eqnarray}
The contribution of Fig.1k+1l is expressed as
\begin{eqnarray} 
\Gamma_{k+l}^{\mu(p)}&=&\frac{{\cal C}(D-F)c_1}{4m_{\Sigma^*}\,f^2}I_{(k+l)K}^{\Sigma\Sigma^*},
\end{eqnarray} 
where
\begin{eqnarray} 
I_{(k+l)K}^{\Sigma\Sigma^*}&=&\bar{u}(p')\tilde F(q)\int\!\frac{d^4 k}{(2\pi)^4}\tilde F^2(k)\slashed{k}\gamma_5\frac{1}
{\slashed{p'}-\slashed{k}-m_{\Sigma^*}}\slashed{q}\gamma_5\frac{1}{\slashed{p}-\slashed{k}-m_{\Sigma^*}}S_{\mu\rho}(p-k)
(k_\rho+z\gamma_\rho\slashed{k})\frac{1}{D_K(k)}u(p)              \nonumber\\
&-&\bar{u}(p')\tilde F(q)\int\!\frac{d^4 k}{(2\pi)^4}\tilde F^2(k)\slashed{k}\gamma_5\frac{1}{\slashed{p'}
-\slashed{k}-m_{\Sigma^*}}\gamma^\mu\gamma_5q_\nu\frac{1}{\slashed{p}-\slashed{k}-m_{\Sigma^*}}S_{\nu\rho}(p-k)
(k_\rho+z\gamma_\rho\slashed{k})\frac{1}{D_K(k)}u(p)              \nonumber\\
&+&\bar{u}(p')\tilde F(q)\int\!\frac{d^4 k}{(2\pi)^4}\tilde F^2(k)(k_\nu+z\slashed{k}\gamma_\nu)\frac{1}{\slashed{p'}
-\slashed{k}-m_{\Sigma^*}}S_{\nu\rho}(p^\prime-k)q_\rho\gamma^\mu\gamma_5\frac{1}{\slashed{p}-\slashed{k}
-m_{\Sigma^*}}\slashed{k}\gamma_5\frac{1}{D_K(k)}u(p)\nonumber\\
&-&\bar{u}(p')\tilde F(q)\int\!\frac{d^4 k}{(2\pi)^4}\tilde F^2(k)(k_\nu+z\slashed{k}\gamma_\nu)\frac{1}{\slashed{p'}
-\slashed{k}-m_{\Sigma^*}}S_{\nu\mu}(p^\prime-k)q_\rho\gamma^\rho\gamma_5\frac{1}{\slashed{p}-\slashed{k}
-m_{\Sigma^*}}\slashed{k}\gamma_5\frac{1}{D_K(k)}u(p).
\end{eqnarray}
The contribution of Fig.~1m+1n is expressed as 
\begin{eqnarray} 
\Gamma_{m+n}^{\mu(p)}&=& \frac{{\cal C}^2}{8f^2}I_{(m+n)K}^{N\Sigma^*},\\
\end{eqnarray}
where
\begin{eqnarray}
I_{(m+n)K}^{N\Sigma^*}&=&\bar{u}(p')\tilde F(q)\int\!\frac{d^4 k}{(2\pi)^4}(k_\sigma+z\slashed{k}
\gamma_\sigma)\tilde F^2(k)\frac{1}{D_K(k)}\frac{1}{\slashed{p'}-\slashed{k}-m_{\Sigma^*}}S_{\sigma\rho}(p^\prime-k)
(g^{\rho\mu}+z\gamma^\rho\gamma^\mu)u(p)                          \nonumber\\
&+&\bar{u}(p')\tilde F(q)\int\!\frac{d^4 k}{(2\pi)^4}(g^{\sigma\mu}+z\gamma^\mu\gamma^\sigma)\tilde F^2(k)
\frac{1}{D_K(k)}\frac{1}{\slashed{p}-\slashed{k}-m_{\Sigma^*}}S_{\sigma\rho}(p-k)(k_\rho+z\gamma_\rho\slashed{k})u(p).
\end{eqnarray}
The  contribution of Fig.~1o+1p is expressed as
\begin{eqnarray} 
\Gamma_{o+p}^{\mu(p)}&=&\frac{{\cal C}^2}{8f^2}I_{(o+p)K}^{N\Sigma^*},\\
\end{eqnarray} 
where
\begin{eqnarray}
I_{(o+p)K}^{N\Sigma^*}&=&-\bar{u}(p')\tilde F(q)\int\!\frac{d^4 k}{(2\pi)^4}(k_\sigma+z\slashed{k}\gamma_\sigma)\tilde F(k)
\frac{1}{D_K(k)}\frac{1}{\slashed{p'}-\slashed{k}-m_{\Sigma^*}}S_{\sigma\rho}(p^\prime-k)
\left(k_\rho-q_\rho+z\gamma_\rho(\slashed{k}-\slashed{q})\right)   \nonumber\\
&\times&\frac{(-2k+q)^\mu}{-2kq+q^2}[F(k-q)-F(k)]u(p)                \nonumber\\
&+&\bar{u}(p')\tilde F(q)\int\!\frac{d^4 k}{(2\pi)^4}\left(k_\sigma+q_\sigma+z(\slashed{k}+\slashed{q})\gamma_\sigma\right)
\frac{(2k+q)^\mu}{2kq+q^2}[F(k+q)-F(k)] \frac{1}{D_K(k)}                              \nonumber\\
&\times&\frac{1}{\slashed{p}-\slashed{k}-m_{\Sigma^*}}S_{\sigma\rho}(p-k)\tilde F(k)(k_\rho+z\gamma_\rho\slashed{k})u(p).
\end{eqnarray}

\providecommand{\href}[2]{#2}\begingroup\raggedright\endgroup

\end{document}